\documentclass[prd,aps,twocolumn,a4paper,showkeys]{revtex4-1}
\usepackage{amsmath}
\usepackage{amsfonts}
\usepackage{amssymb}	
\usepackage{graphicx}
\usepackage{bm}
\usepackage{color}
\usepackage{commath}
\allowdisplaybreaks

\def\GMc2{G M_{\odot} c^{-2}}

\def\lm{{\ell m}}

\def\lm{{\ell m}}

\def\de{\partial}
\def\lm{{\ell m}}

\def\ii{{\rm i}}

\def\F{{\cal F}}

\newcommand\be{\begin{equation}}
\newcommand\ee{\end{equation}}

\def\TEOBResumShm{\texttt{TEOBiResumS\_SM}}

\begin{document}
\title{Faithful analytical effective one body waveform model for spin-aligned,
       moderately eccentric, coalescing black hole binaries}
\author{Danilo \surname{Chiaramello}${}^{1,2}$}
\author{Alessandro \surname{Nagar}${}^{2,3}$}
\affiliation{${}^{1}$ Dipartimento di Fisica, Universit\`a di Torino, via P. Giuria 1, 10125 Torino, Italy}
\affiliation{${}^2$INFN Sezione di Torino, Via P. Giuria 1, 10125 Torino, Italy}
\affiliation{${}^3$Institut des Hautes Etudes Scientifiques, 91440 Bures-sur-Yvette, France}

\begin{abstract}
 We present a new effective-one-body (EOB) model for eccentric binary coalescences.
 The model stems from the state-of-the-art model \TEOBResumShm{} for circularized coalescing 
 black-hole binaries, that is modified to explicitly incorporate eccentricity effects {\it both}
 in the radiation reaction and in the waveform. Using Regge-Wheeler-Zerilli type calculations of
 the gravitational wave losses as benchmarks, we find that a rather accurate ($\sim 1\%$) 
 expression for the radiation reaction  along mildly eccentric orbits ($e \sim 0.3$) is given 
 by {\it dressing} the current, EOB-resummed, circularized angular momentum flux, 
 with a leading-order  (Newtonian-like) prefactor valid along general orbits. An analogous 
 approach is implemented for the waveform multipoles. The model is then completed by 
 the usual merger-ringdown part informed by circularized numerical relativity (NR) simulations. 
 The model is validated against the 22, publicly available, NR simulations calculated by
 the Simulating eXtreme Spacetime (SXS) collaboration, with mild eccentricities,
 mass ratios between 1 and 3 and up to rather large dimensionless spin values ($\pm 0.7$).
 The maximum EOB/NR unfaithfulness, calculated with Advanced LIGO noise,
 is at most of order $3\%$. The analytical framework presented here should be seen
 as a promising starting point for developing highly-faithful waveform templates
 driven  by eccentric dynamics for present, and possibly future, gravitational wave detectors.
   \end{abstract}
   
\date{\today}

\maketitle

\section{Introduction}
\label{sec:intro}
Parameter estimates of all gravitational wave (GW) signals from coalescing
binaries are done under the assumption that the inspiral is quasi-circular~\cite{LIGOScientific:2018mvr}.
This is motivated by the efficient circularization of the inspiral due
to gravitational wave emission. In addition, no explicit evidence for
eccentricity for some events was found~\cite{Abbott:2016wiq,Nitz:2019spj,Romero-Shaw:2020aaj}.
However, recent population synthesis studies~\cite{Samsing:2013kua,Rodriguez:2016kxx,Belczynski:2016obo,Samsing:2017xmd} 
suggest that active galactic nuclei and globular clusters may host a population of eccentric binaries.
Currently, there are no  ready-to-use waveform models  that accurately
combine both eccentricity and spin effects over the entire parameter space.
Recently, numerical relativity (NR) started producing
surveys of eccentric, spinning binary black hole (BBH) coalescence
waveforms~\cite{Hinder:2017sxy,Ramos-Buades:2019uvh,Huerta:2019oxn},
and a NR-surrogate waveform model for nonspinning eccentric binaries up to
mass ratio $q=10$ exists~\cite{Huerta:2019oxn}. On the analytical side,
Refs.~\cite{Klein:2018ybm,Tiwari:2019jtz} provided closed-form
eccentric inspiral templates  (based on the Quasi-Keplerian approximation).
Similarly, a few exploratory effective-one-body 
(EOB)-based~\cite{Buonanno:1998gg,Buonanno:2000ef,Buonanno:2005xu,Damour:2015isa}   
studies were recently performed~\cite{Cao:2017ndf,Hinderer:2017jcs,Liu:2019jpg}.
In particular Refs.~\cite{Cao:2017ndf,Liu:2019jpg} introduced and tested 
{\tt SEOBNRE}, a way to incorporate eccentricity within the 
{\tt SEOBNRv1}~\cite{Taracchini:2012ig} circularized waveform model. 
However, the {\tt SEOBNRv1} model is outdated now, since it does not accurately
cover high-spins, nor mass ratios up to 10. This drawback is inherited
by the {\tt SEOBNRE} model~\cite{Liu:2019jpg}.

In this article, we modify a highly NR-faithful EOB multipolar
waveform model for circularized coalescing BBHs, \TEOBResumShm{}~\cite{Nagar:2018zoe,Nagar:2020pcj},
to incorporate eccentricity-dependent effects.
The EOB formalism relies on three building blocks: (i) a Hamiltonian, that describes the
conservative part of the relative dynamics; (ii) a radiation reaction force,
that accounts for the back-reaction onto the system due to the GW losses
of energy and angular momentum; (iii) a prescription for computing
the waveform. Including eccentricity requires modifications to blocks (ii) and (iii)
with respect to the quasi-circular case.
%We do this by replacing some leading (Newtonian level) circularized expressions
%with their counterparts along generic orbits, and by adding some more sophisticated
%analytical results at (resummed) 2PN accuracy in the radial
%radiation reaction~\cite{Bini:2012ji}.

\section{Radiation reaction and waveform for eccentric inspirals}
\label{sec:RR}
Within the EOB formalism, we use phase-space variables $(r,\varphi,p_\varphi,p_{r_*})$,
related to the physical ones by $r=R/(GM)$ (relative separation),
$p_{r_*}=P_{R_*}/\mu$ (radial momentum), $p_\varphi=P_\varphi/(\mu GM)$ (angular momentum)
and $t=T/(GM)$ (time),
where $\mu\equiv m_1 m_2/M$ and $M=m_1+m_2$. The radial momentum is $p_{r_*}\equiv (A/B)^{1/2}p_r$,
where $A$ and $B$ are the EOB potentials. The EOB Hamiltonian is
$\hat{H}_{\rm EOB}\equiv H_{\rm EOB}/\mu=\nu^{-1}\sqrt{1+2\nu(\hat{H}_{\rm eff}-1)}$, with 
$\nu\equiv \mu/M$ and $\hat{H}_{\rm eff}=\tilde{G}p_\varphi + \hat{H}^{\rm orb}_{\rm eff}$, 
where $\tilde{G}p_\varphi$ incorporates odd-in-spin (spin-orbit) effects while 
$\hat{H}^{\rm orb}_{\rm eff}$ incorporates even-in-spin effects~\cite{Nagar:2018zoe}. We denote
dimensionless spin variables as $\chi_{i}\equiv S_i/m_i^2$.
The \TEOBResumShm{}~\cite{Nagar:2016ayt,Messina:2018ghh,Nagar:2020pcj}
waveform model is currently the most NR faithful model versus
the {\tt zero\_det\_highP} Advanced LIGO design sensitivity~\cite{dcc:2974}).
Reference~\cite{Nagar:2020pcj} found that the maximum value of the EOB/NR unfaithfulness
is always below $0.5\%$ all\footnote{Modulo a single outlier at $0.7\%$}
over the current release of the SXS NR waveform
catalog~\cite{Chu:2009md,Lovelace:2010ne,Lovelace:2011nu,Buchman:2012dw,
Hemberger:2013hsa,Scheel:2014ina,Blackman:2015pia,
Lovelace:2014twa,Mroue:2013xna,Kumar:2015tha,Chu:2015kft,
Boyle:2019kee,SXS:catalog}. This is achieved by NR informing a
4.5PN spin-orbit effective function $c_3(\nu,\chi_1,\chi_2)$
and an effective 5PN function $a_6^c(\nu)$ entering the
Pad\'e resummed radial potential $A(r)$.
[see Eqs.~(39) and~(33) of Ref.~\cite{Nagar:2019wds}].
The two Hamilton's equations that take account of GW losses are
\begin{align}
  \dot{p}_\varphi &= \hat{\F}_\varphi, \\
  \dot{p}_{r_*}   &= \sqrt{\frac{A}{B}} \left(-\de_{r} \hat{H}_{\rm EOB} + \hat{\F}_r \right),
\end{align}
where $(\hat{\F}_\varphi,\hat{\F}_r)$ are the two radiation reaction forces.
In the quasi-circular case~\cite{Nagar:2018zoe,Nagar:2019HM} one sets
$\hat{\F}_r=0$. Here, we use $\hat{\F}_r\neq 0$ and $\hat{\F}_\varphi$
explicitly includes noncircular terms. The main technical issue 
is to build (resummed) expressions of $(\hat{\F}_\varphi,\hat{\F}_r)$
that are reliable and robust up to merger. Building
upon Ref.~\cite{Gopakumar:1997ng}, Ref.~\cite{Bini:2012ji}
derived the 2PN-accurate, generic expressions of
$(\hat{\F}_\varphi,\hat{\F}_r)$, which are unsuited to
drive the transition from the EOB inspiral to plunge
and merger: they are nonresummed and generally
unreliable in the strong-field regime (see below).
The forces are related to the instantaneous losses of energy and angular 
momentum through GWs. Following Ref.~\cite{Bini:2012ji},
there exists a gauge choice such that the balance equations read
\begin{align}
  -\dot{J}^\infty &=  \F_\varphi,\\ 
  -\dot{E}^\infty&=\dot{r}\F_r + \dot{\varphi}\F_\varphi + \dot{E}_{\rm Schott},
\end{align}
%=============================
\begin{figure}[t]
	\center
	\includegraphics[width=0.45\textwidth]{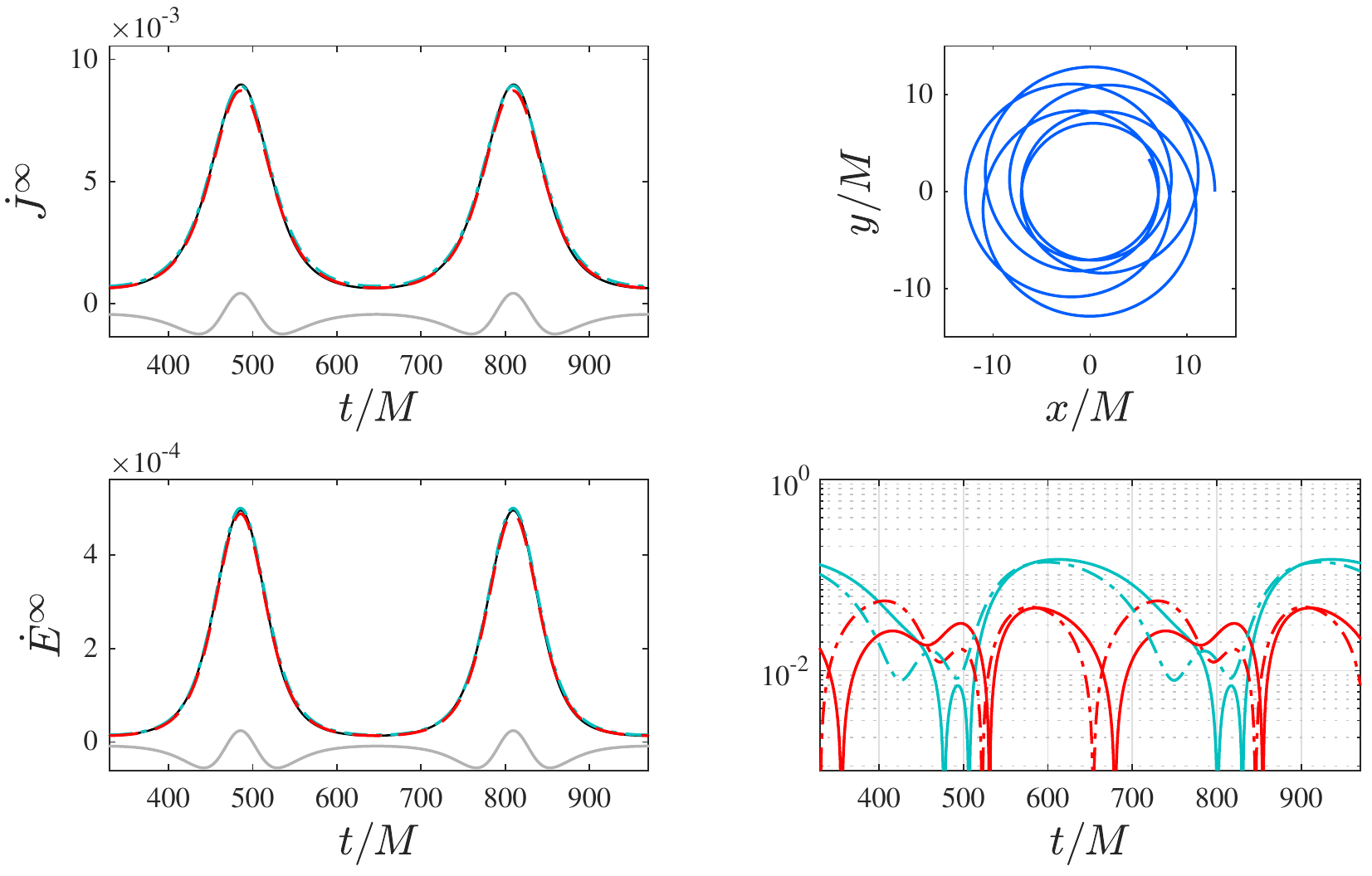}	
	\caption{\label{fig:geo}Test-particle orbiting a Schwarzschild black hole
          with semilatus rectum $p=9$ and eccentricity $e=0.3$. Different analytical
          representations of the angular momentum and energy fluxes
           $(\dot{J}^\infty, \dot{E}^{\infty})$ are compared with the numerical ones 
           using the Regge-Wheeler-Zerilli (RWZ) formalism (black).
           Left panels: the nonresummed 2PN-accurate ones of Ref.~\cite{Bini:2012ji},
           (gray); the resummed version via Eq.~\eqref{eq:BDnc} (light-blue);
           the resummed version using the noncircular Newtonian prefactor
           (red), Eq.~\eqref{eq:EOB-newt-nc}. The relative differences
           in $(\dot{J}^{\infty},\dot{E}^{\infty})$
           are shown in the bottom-right panel: $|\delta \dot{J}^{\infty}/\dot{J}_{\rm RWZ}^{\infty}|$
           (solid) and $|\delta \dot{E}^{\infty} / \dot{E}_{\rm RWZ}^{\infty}|$ (dashed),
           color scheme as above. On average, Eq.~\eqref{eq:EOB-newt-nc}
           delivers the closest analytical/numerical agreement.}
	\center
\end{figure}
%=============================
where $E_{\rm Schott}$ is the Schott energy 
(see Ref.~\cite{Damour:2011fu,Bini:2012ji} and references therein), 
$(\dot{E},\dot{J})^{\infty}$ are the energy and angular 
momentum fluxes at infinity, while $\F_{\varphi,r}\equiv\mu\hat{\F}_{\varphi,r}$.
To build the resummed expressions of the functions $(\F_\varphi,\F_r,\dot{E}_{\rm Schott})$
and evaluate their strong-field reliability, we adopt the
procedure that proved fruitful in the circularized
case~\cite{Damour:1997ub,Damour:2008gu,Pan:2010hz,Nagar:2016ayt,Messina:2018ghh,Nagar:2019wrt}:
any analytical choice for $(\F_\varphi,\F_r,\dot{E}_{\rm Schott})$
is tested by comparisons with the energy and angular momentum fluxes
emitted by a test particle orbiting a Schwarzschild black
hole on eccentric orbits.
We focus first on $\F_\varphi$. We start with the
2PN-accurate result of Ref.~\cite{Bini:2012ji}
[see Eq.~(3.70) and Appendix~D therein], $\F_\varphi^{\rm 2PN} (r, p_{r}, p_{\varphi})$, 
reexpress it in terms of $p_{r_*}$, and factor it in a
circular part (defined imposing $p_{r_*} = \dot{p}_{r_*} = 0$),
$\F^{\rm 2PN_{c}}_{\varphi} (r)$, and a noncircular contribution,
$\F^{\rm 2PN_{\rm nc}}_{\varphi} (r, p_{r_*}, p_\varphi)$, so that
$\F^{\rm 2PN}_\varphi(r, p_{r_*},p_\varphi)= \F^{\rm 2PN_{c}}_{\varphi}(r)\F^{\rm 2PN_{\rm nc}}_{\varphi}(r,p_{r_*},p_\varphi)$.
A route to improve the strong-field behavior of this expression is
to replace $\F^{\rm 2PN_{c}}_{\varphi}(r)$ with the corresponding EOB-resummed
expression~\cite{Damour:2008gu}
(notably, in its latest avatar~\cite{Nagar:2020pcj,Nagar:2016ayt,Messina:2018ghh}).
To do so, the radial EOB coordinate $r$ in  $\F^{\rm 2PN_{c}}_{\varphi}(r)$
is first replaced by the circularized frequency variable $x\equiv \Omega_{\rm circ}^{2/3}$,
Eq.~(5.22) of Ref.~\cite{Bini:2012ji} at 2PN accuracy; then this 2PN-accurate expression
is replaced by $\F_\varphi^{\rm EOB_c}(x)=-32/5\nu^2 x^{7/2}\hat{f}(x)$, where
$\hat{f}\equiv \left(F_{22}^{\rm Newt}\right)^{-1}\sum_\lm F_{\lm}$ is the
factored flux function~\cite{Damour:2008gu}, with all multipoles  (except $m=0$ ones)
up to $\ell=8$. Finally, the function $\F_\varphi^{\rm EOB_c}(x)$ is computed along
the noncircular dynamics. We do so by using the circular frequency
$\Omega_{\rm circ}\equiv \de_{p_\varphi}\hat{H}_{\rm EOB}|_{p_\varphi=j,p_{r_*}=0}$,
where $j^2\equiv -A'(u)/(u^2A(u))'$ is the (squared) circular angular momentum,
$u\equiv r^{-1}$ and $(\cdot)'\equiv \de_u$. Note that in the resummed flux,
we use $\{p_\varphi,\hat{H}_{\rm EOB}(r,p_{r_*},p_\varphi),\hat{H}_{\rm eff}(r,p_{r_*},p_{\varphi})\}$
computed along the general dynamics. The 2PN-accurate noncircular contribution
$\F^{\rm 2PN_{nc}}_{\varphi} \equiv f_{\varphi}^{N_{\rm nc}} + c^{-2} f_{\varphi}^{\rm 1PN_{nc}} + c^{-4} f_{\varphi}^{\rm 2PN_{\rm nc}}$
is resummed using a $(0,2)$ Pad\'e approximant. We have
\be
\label{eq:BDnc}
\F_\varphi^{\rm EOB_{2PN_{nc}}}\equiv\F_\varphi^{\rm EOB_c}(x(r))P^0_2[\F_\varphi^{\rm 2PN_{nc}}(r,p_{r_*},p_{\varphi})].
\ee
%==============
% RWZ insplunge
%==============
\begin{figure}[t]
\center
\includegraphics[width=0.45\textwidth]{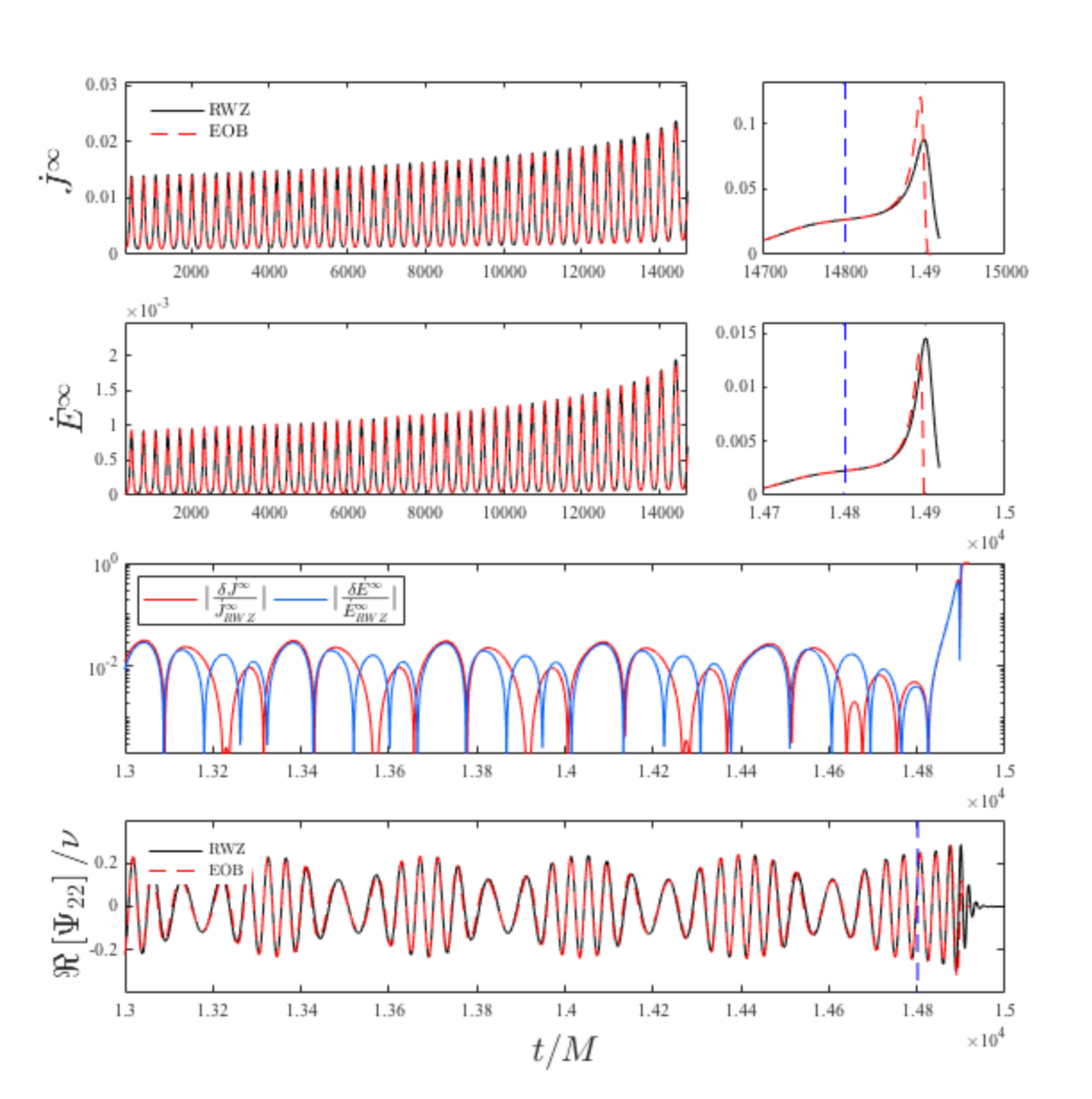}
\caption{\label{fig:insplunge} Test particle (with mass ratio $\mu/M=10^{-3}$, $e=0.3)$ and $p=8$)
  plunging over a Schwarzschild black hole. EOB/RWZ comparison between
  energy and angular momentum fluxes (top three panels) and waveforms
  (bottom panel). Vertical line: crossing of the stability threshold ($p = 6 + 2 e$) and beginning of the plunge.}
\center
\end{figure}
%==============
Alternatively, we recall that the force used to drive
the EOB quasi-circular inspiral is
\begin{equation}
  \label{eq:EOB-circ}
\F_\varphi^{\rm EOB_{qc}}=-\dfrac{32}{5}\nu^2 r_\omega^4\Omega^5\hat{f}(\Omega),
\end{equation}
where $\Omega\equiv \dot{\varphi}$, which yields a more faithful
representation of GW losses during the plunge~\cite{Damour:2006tr,Damour:2007xr}.
This expression is the leading quasi-circular term of the Newtonian
angular momentum flux, obtained from Eq.~(3.26) of Ref.~\cite{Bini:2012ji},
neglecting higher-order derivatives of $(r,\Omega)$. We can thus improve
Eq.~\eqref{eq:EOB-circ} multiplying it with the Newtonian noncircular factor
\begin{align}
\label{eq:Fnewt}
&\hat{f}_\varphi^{\rm {Newt}_{nc}}=1+\dfrac{3}{4}\dfrac{\ddot{r}^2}{r^2\Omega^4} - \dfrac{\ddot{\Omega}}{4\Omega^3}
+\dfrac{3\dot{r}\dot{\Omega}}{r\Omega^3} \\
& +\dfrac{4\dot{r}^2}{r^2\Omega^2}
+\dfrac{\ddot{\Omega}\dot{r}^2}{8 r^2\Omega^5} + \dfrac{3}{4}\dfrac{\dot{r}^3\dot{\Omega}}{ r^3\Omega^5}
+\dfrac{3}{4}\dfrac{\dot{r}^4}{r^4\Omega^4} + \dfrac{3}{4}\dfrac{\dot{\Omega}^2}{\Omega^4}\nonumber\\
&-\dddot{r}\left(\dfrac{\dot{r}}{2r^2\Omega^4}+\dfrac{\dot{\Omega}}{8r\Omega^5}\right)%\nonumber\\
+\ddot{r}\left(-\dfrac{2}{r\Omega^2}+\dfrac{\ddot{\Omega}}{8r \Omega^5}+\dfrac{3}{8}\dfrac{\dot{r}\dot{\Omega}}{r^2\Omega^5}\right)\nonumber,
\end{align}
in order to get
\begin{equation}
  \label{eq:EOB-newt-nc}
\F_\varphi^{\rm EOB_{Newt_{nc}}}=-\dfrac{32}{5}\nu^2 r_\omega^4\Omega^5\hat{f}_\varphi^{\rm {Newt}_{nc}}\hat{f}(\Omega).
\end{equation}
Although this expression incorporates formally
{\it less} noncircular PN information than
Eq.~\eqref{eq:BDnc}, the time-derivatives
(and $\hat{f}(\Omega)$ as well) are obtained
from the full EOB (resummed) equations of motion
rather the 2PN ones used in $\F_\varphi^{\rm 2PN_{nc}}$.
For $\F_r$, we build on Ref.~\cite{Bini:2012ji} and we use
$\F_r = 32/3p_{r_*}/r^4P^0_2[\hat{\F}^{\rm 2PN}_r]$,
where $P^0_2$ is the  $(0,2)$ Pad\'e approximant and 
$\hat\F^{\rm 2PN}_r=f_r^N + c^{-2}f_r^{\rm 1PN}+c^{-4}f_r^{\rm 2PN}$
is the 2PN accurate expression calculated from Eqs.~(3.70) and (D9-D11) of Ref.~\cite{Bini:2012ji}.
We adopt an analogous approach to deal with the Schott energy, as given by
Eqs.~(3.57) and (C1-C4) of Ref.~\cite{Bini:2012ji}. We factorize it in  
circular and noncircular parts that are both resummed with the  $P^0_2$
Pad\'e approximant, so to have
$E_{\rm Schott}=16/5 p_{r_*}/r^3 P^0_2[E_{\rm Schott}^{\rm c}] P^0_2[E_{\rm Schott}^{\rm nc}]$, 
where $E_{\rm Schott}^{\rm nc} = E^{\rm nc,0}_{\rm Schott} + c^{-2} E^{\rm nc, 1PN}_{\rm Schott}
+ c^{-4} E^{\rm nc,2PN}_{\rm Schott}$.
%
%--------------------------
% SXS eccentric simulations
%--------------------------
\begin{table}[t]
   \caption{\label{tab:SXS} SXS simulations with eccentricity analyzed in this work. From left to right: the 
   ID of the simulation; the mass ratio $q\equiv m_1/m_2\geq 1$; the individual spins $(\chi_1,\chi_2)$;
   the estimated NR eccentricity at first apastron $e_{\rm \omega}^{\rm NR}$; the initial EOB eccentricity
   $e^{\rm EOB}$ and apastron frequency $\omega_{a}^{\rm EOB}$ which allow us to get a good EOB/NR frequency agreement.}
   \begin{center}
     \begin{ruledtabular}
\begin{tabular}{ c c c c c |l l| c} 
  SXS & $q$ & $\chi_1$ & $\chi_2$ & $e^{\rm NR}_{\rm \omega}$ & $e^{\rm EOB}$ & $\omega_{a}^{\rm EOB}$ & $\max(\bar{F})[\%]$ \\
  \hline
  \hline
1355 & 1 & 0 & 0 & 0.062  & 0.089   & 0.0280475  &  1.30\\
1356 & 1 & 0 & 0 & 0.102 & 0.1503  & 0.019077  &  1.03\\
1359 & 1 & 0 & 0 & 0.112 & 0.18    & 0.021495 & 1.22 \\
1357 & 1 & 0 & 0 & 0.114 & 0.1916   & 0.019617 & 1.20\\
1361 & 1 & 0 & 0 & 0.160 & 0.23437 & 0.02104   & 1.56\\
1360 & 1 & 0 & 0 & 0.161 & 0.2415   & 0.019635   & 1.52\\
1362 & 1 & 0 & 0 & 0.217  & 0.30041   & 0.0192 & 0.89\\
1364 & 2 & 0 & 0 & 0.049 & 0.0843   & 0.025241   & 0.86\\
1365 & 2 & 0 & 0 & 0.067 & 0.11    & 0.023987 &1.00\\
1367 & 2 & 0 & 0 & 0.105 & 0.1494  & 0.026078    &0.92\\
1369 & 2 & 0 & 0 & 0.201 & 0.309    & 0.01755  &1.38\\
1371 & 3 & 0 & 0 & 0.063 & 0.0913   & 0.029058   &0.57\\
1372 & 3 & 0 & 0 & 0.107  & 0.149    & 0.026070 &0.95\\
1374 & 3 & 0 & 0 & 0.208 & 0.31405 & 0.016946   &0.78\\
\hline
89   & 1    & $-0.5$  & 0        & 0.047  & 0.071   & 0.0178279  & 0.96  \\
1136 & 1    & $-0.75$ & $-0.75$  & 0.078  & 0.121    & 0.02728 & 0.58  \\
321  & 1.22 & $+0.33$ & $-0.44$  & 0.048  & 0.076    & 0.02694 &1.47  \\
322  & 1.22 & $+0.33$ & $-0.44$  & 0.063  & 0.0984    & 0.026895 &1.18  \\
323  & 1.22 & $+0.33$ & $-0.44$  & 0.104  & 0.141     & 0.025965 &1.57 \\
324  & 1.22 & $+0.33$ & $-0.44$  & 0.205  & 0.2915     & 0.019067 & 2.25\\
1149 & 3    & $+0.70$ & $+0.60$   & 0.037  & $0.0617$ & $0.0266802$ & 3.16\\
1169 & 3    & $-0.70$ & $-0.60$   & 0.036  & $0.049$ & $0.024285$ & 0.17      % 

 \end{tabular}
 \end{ruledtabular}
 \end{center}
 \end{table}
%===================
Equations~\eqref{eq:BDnc}-\eqref{eq:EOB-newt-nc} are specialized to the
test particle limit ($\nu=0$) and computed along the eccentric,
conservative, dynamics of a particle orbiting a Schwarzschild black hole.
The result is compared with the fluxes computed using Regge-Wheeler-Zerilli
(RWZ) black hole perturbation theory~\cite{Regge:1957td,Zerilli:1970se,Nagar:2005ea}. 
To accurately extract waves at future null infinity, we adopt the
hyperboloidal layer method of~\cite{Bernuzzi:2011aj} and compute the
fluxes with the usual expressions\footnote{We removed the (negligible)
  $m=0$ modes since their analytic representation is poor.}
of Ref.~\cite{Nagar:2005ea}, including all multipoles up to $\ell=8$.
Figure~\ref{fig:geo} shows the illustrative case of an orbit with
semilatus rectum $p=9$ and eccentricity $e=0.3$. 
The apastron is $r_1=p/(1-e)$, and the periastron is
$r_2=p/(1+e)$~\footnote{Semilatus rectum and eccentricity are \textit{defined} by their relationships with 
the apastron and periastron radii. Explicit formulae relating $\left( E, p_{\varphi} \right)$ 
and $\left( p, e \right)$, which we use to set the initial data 
for an orbit specified by its eccentricity and semilatus rectum, 
as well as to compute $p$ and $e$ at each step of the motion, 
can be derived by solving the equations: $E = \hat{H}_{\rm EOB} \left( r_{1,2}, p_{\varphi}, 
p_{r_*} = 0 \right)$; finding, e.g., in the Schwarzschild case:
$E^{2} = ((p - 2)^2 - 4 e^2)/(p \left( p - 3 - e^2 \right))$
and  $p_{\varphi}^2 = p^2/(p - 3 - e^2)$.
}.
The figure indicates that $\F_\varphi^{\rm EOB_{Newt_{nc}}}$ 
delivers analytical energy and angular momentum fluxes (red lines)
that are, on average, in better agreement with the RWZ ones than
those obtained from $\F_\varphi^{\rm EOB_{2PN_{nc}}}$, which increase
up to a $10\%$ fractional difference at apastron. We adopt then
$\F_\varphi^{\rm EOB_{Newt_{nc}}}$ as analytical representation
of the angular momentum flux along generic orbits. 
The maximal analytical/RWZ flux relative differences are $\sim 10^{-2}$.
The robustness of this result is checked by considering several orbits
with $p$ varying from just above the stability threshold ($p = 6 + 2 e$) up to $p = 21$, 
and for each $p$ we consider $0 \leq e \leq 0.9$. We then compute
the relative flux differences $(\delta\dot{E},\delta\dot{J})$ at periastron 
for each $(p,e)$. We find that, for each value of 
$p$, $(\delta\dot{E},\delta\dot{J})$ are at most of the order 
of $10\%$ for $e=0.9$. More interestingly if $e\lesssim 0.3$, 
the fractional differences do not exceed the $5\%$ level.
%
%
%====================
% Time-domain phasing
%===================
\begin{figure}[t]
\center
\includegraphics[width=0.45\textwidth]{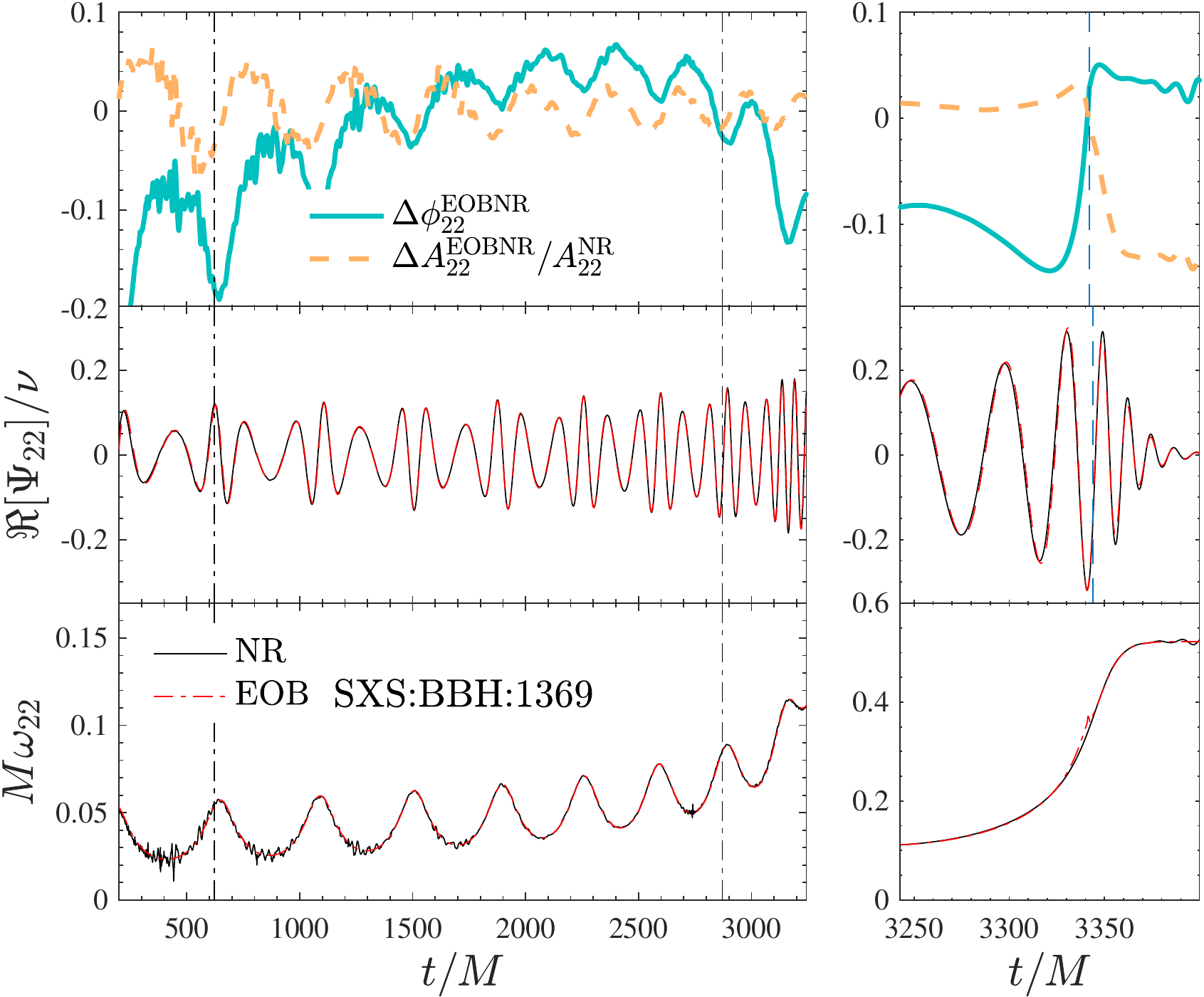}
\caption{\label{fig:h22}Illustrative EOB/NR time-domain waveform comparison for the nonspinning
  configuration {\tt SXS:BBH:1369}, with $e_\omega^{\rm NR}=0.201$, $q=2$ and $\chi_1=\chi_2=0$.}
\end{figure}
Let us consider now the waveform emitted from the transition from
inspiral to plunge, merger and ringdown as driven by $(\F_r,\F_\varphi^{\rm EOB_{Newt_{nc}}})$,
focusing on a test-particle (of mass ratio $\mu/M=10^{-3}$)
on a Schwarzschild background. To efficiently compute, along the relative dynamics,
up to the third time-derivative  of the phase-space variables entering
Eq.~\eqref{eq:Fnewt}, we suitably generalize the iterative analytical 
procedure used in Appendix~A of Ref.~\cite{Damour:2012ky} to calculate $\ddot{r}$. 
We checked that two iterations are sufficient to obtain an excellent approximation
$(\simeq 10^{-3})$ of the derivatives computed numerically. 
An illustrative waveform is displayed in Fig.~\ref{fig:insplunge} for $p=8$
and $e=0.3$ (initial values). The top three rows of the figure highlight the numerical
consistency ($\sim 10^{-2}$) between the RWZ  angular momentum and energy
fluxes and their analytical counterparts. The corresponding waveform 
is shown (in black) in the fourth row of the plot. The gravitational waveform
is decomposed in multipoles as
$h_+-\ii h_\times=D_L^{-1}\sum_\lm h_\lm {}_{-2}Y_{\lm}$, where $D_L$
is the luminosity distance and ${}_{-2}Y_{\lm}$ the $s=-2$ spin-weighted
spherical harmonics. We use below the RWZ normalized 
variable $\Psi_{\lm}=h_{\lm}/\sqrt{(\ell+2)(\ell+1)\ell(\ell-1)}$.
%===============
% Unfaithfulness
%===============
\begin{figure}[t]
\center
\includegraphics[width=0.45\textwidth]{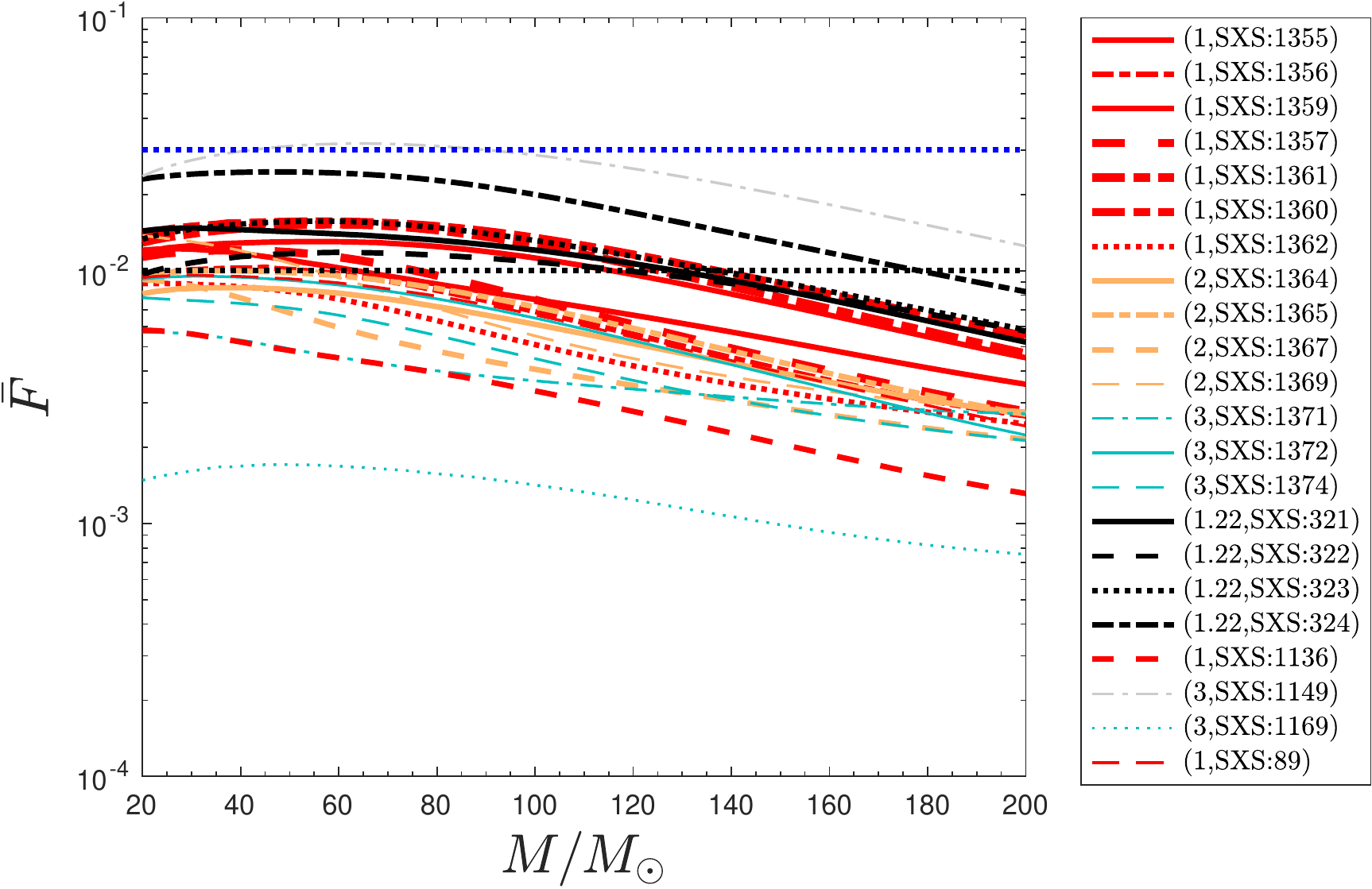}
\caption{\label{fig:barF}EOB/NR unfaithfulness computed over the eccentric SXS
  simulations publicly available. The horizontal lines mark
  the $3\%$ and $1\%$ values.}
\end{figure}
%
%============================================================
% EOB/NR comparison for (3,3) and (4,4) mode for SXS:BBH:1369
%============================================================
\begin{figure}[t]
\center
\includegraphics[width=0.45\textwidth]{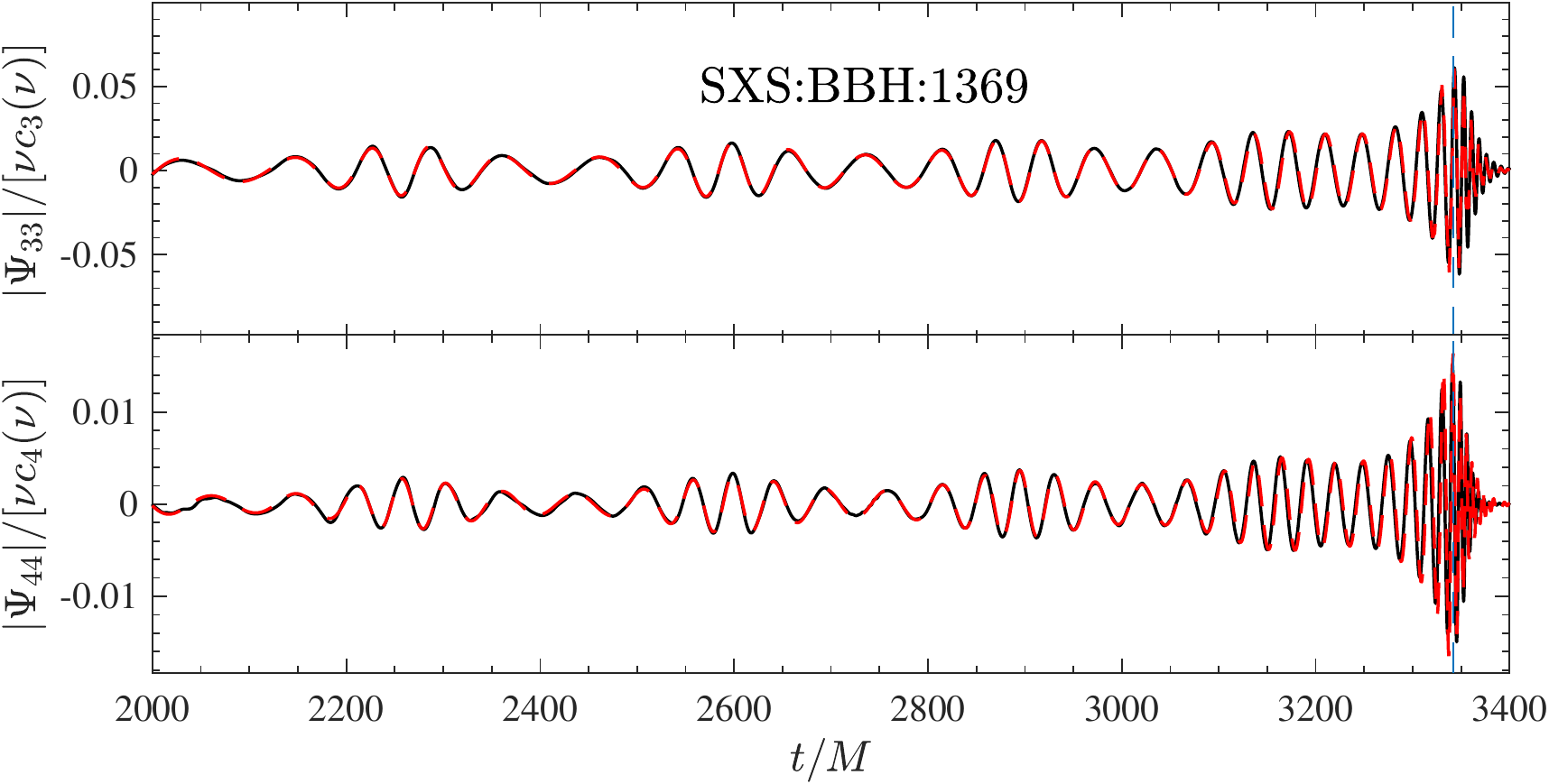}
\caption{\label{fig:hm}Illustrative EOB/NR comparison for {\tt SXS:BBH:1369}
  for the $(3,3)$ and $(4,4)$ waveform modes. Normalization
  constants are $c_3=\sqrt{1-4\nu}$ and $c_4=1-3\nu$.}
\end{figure}
A detailed analysis of the properties of the RWZ waveform, such as the
excitation of QNMs, etc., will be presented elsewhere.
Here we employ it as a
{\it target}, an ``exact'' waveform to validate the EOB one.
Within the EOB formalism, each multipole is factorized as $h_{\lm}=h_\lm^{(N,\epsilon)}\hat{h}_\lm$,
where $h_\lm^{(N,\epsilon)}$ is the Newtonian (leading-order) prefactor, $\hat{h}_\lm$ is
the resummed relativistic correction~\cite{Damour:2008gu} and $\epsilon$ the parity of $\ell+m$.
The circularized prefactor $h_\lm^{(N,\epsilon)}$ is replaced by its
general expression obtained computing the time-derivatives of the
Newtonian mass and current multipoles. We have $h_\lm^{(N,0)}\propto e^{\ii m\varphi} I^{(\ell)}_\lm$
and  $h_\lm^{(N,1)}\propto e^{\ii m \varphi} S^{(\ell)}_\lm$,
where $(\ell)$ indicates the $\ell$-th time-derivative
and $I_\lm\equiv r^\ell e^{-\ii m\varphi}$ and
$S_\lm\equiv r^{\ell +1}\Omega e^{-\ii m\varphi}$ are the Newtonian 
mass and current multipoles. The $\ell=m=2$ mode of the
analytical waveform is superposed, as a red line, in the bottom row
of Fig.~\ref{fig:insplunge}, showing excellent agreement with the RWZ
one essentially up to merger\footnote{The analytical waveform is
not completed with any RWZ-informed representation of merger and postmerger
phase.}. A similar agreement is found for subdominant modes.

\section{Comparison with numerical relativity simulations}
The complete, $\nu$-dependent, radiation reaction of above replaces
now the standard one used in  \TEOBResumShm{}, so to consistently drive
an eccentric inspiral. Everything is analogous to the test-particle case,
aside from (i) the initial conditions at the apastron, which are more involved because
of the presence of spin, though they are a straightforward generalization 
of those of~\cite{Hinderer:2017jcs}; (ii) similar complications for the 
time-derivatives needed in $\hat{f}^{{\rm Newt}_{\rm nc}}$. The EOB waveform with
the noncircular Newtonian prefactors is completed by next-to-quasi-circular
(NQC) corrections and the NR-informed circularized ringdown~\cite{Nagar:2018zoe,Nagar:2020pcj}.
Differently from the circularized case, the NQC correction factor is smoothly
activated in time just when getting very close to merger, so to avoid
spurious contaminations during the inspiral. Also, no iteration on the NQC amplitude
parameters is performed~\cite{Nagar:2020pcj}.
We assess the quality of the analytic waveforms by comparing
them with the sample of eccentric NR simulations publicly available in the
SXS catalog~\cite{Chu:2009md,Lovelace:2010ne,Lovelace:2011nu,Buchman:2012dw,
Hemberger:2013hsa,Scheel:2014ina,Blackman:2015pia,
Lovelace:2014twa,Mroue:2013xna,Kumar:2015tha,Chu:2015kft,
Boyle:2019kee,SXS:catalog} that are listed in Table~\ref{tab:SXS}.
We carry out both time-domain comparisons and compute the EOB/NR unfaithfulness.
To do so correctly  the EOB evolution should be started in such a way that the
eccentricity-induced frequency oscillations are consistent with the corresponding
ones in the NR simulations. Since the eccentricities are gauge dependent,
their nominal values are meaningless for this purpose. EOB and NR waveforms
are then aligned in the time-domain~\cite{Damour:2012ky} during the early inspiral
and then we progressively vary the initial GW frequency at apastron, $\omega_a^{\rm EOB}$, 
and eccentricity, $e^{\rm EOB}$, until we achieve minimal fractional differences
($\simeq 10^{-2}$) between the EOB and NR GW frequencies. To facilitate the
parameter choice, we also estimate the initial 
(at first apastron) eccentricity of each NR simulation, $e_{\omega}^{\rm NR}$, 
using the method proposed in Eq.~(2.8) of Ref.~\cite{Ramos-Buades:2019uvh}, 
where $e_\omega$ is deduced from the frequency oscillations; we here employ,
however, the frequency of the $(2,2)$ mode, as opposed to the orbital
frequency as done in Ref.~\cite{Ramos-Buades:2019uvh}. The last two columns
of Table~\ref{tab:SXS} contain the values of $(\omega_a^{\rm EOB},e^{\rm EOB})$
that lead to the best agreement between NR and EOB waveforms.
An illustrative time-domain comparison, for  {\tt SXS:BBH:1369} is shown in
Fig.~\ref{fig:h22}. Figure~\ref{fig:barF} shows the EOB/NR unfaithfulness
$\bar{F}\equiv 1-F$ (see Eq.~(48) of Ref.~\cite{Nagar:2020pcj})
computed with the {\tt zero\_det\_highP}~\cite{dcc:2974} Advanced-LIGO power spectral density.
Both NR and EOB waveforms (starting at approximately the same frequency)
were suitably tapered in the early inspiral. From Table~\ref{tab:SXS},
$\max(\bar{F})$ is always comfortably below $3\%$ except for the,
small-eccentricity, dataset {\tt SXS:BBH:1149}, with $\max(\bar F)=3.16$.
We believe that this is the effect of the suboptimal choice of $(a_6^c,c_3)$
(see below) and is {\it not} related to the  modelization of eccentricity effects.
By contrast, for large eccentricities, the $\bar{F}$ computation may be
influenced by the accuracy of NR simulations, which get progressively
more noisy increasing $e_\omega$ (see e.g. bottom panel of Fig.~\ref{fig:h22};
a similar behavior is also found for {\tt SXS:BBH:324}).
The accumulated phase difference at meger
(always $\sim 1$~rad) is mostly due to the previously determined~\cite{Nagar:2020pcj}
$(a_6^c,c_3)$ values, that depend on the circularized waveform
and radiation reaction. Consistently, when our generalized framework
is applied to circularized (nonspinning) binaries, we find that
$\max(\bar{F})$ varies between $1.25\%$ ($q=1$) and $0.21\%$ ($q=8$).
These values are about one order of magnitude {\it larger} than
those of \TEOBResumShm{}(see Fig.~13 of~\cite{Nagar:2019wds}).
Forthcoming work will present a retuning of $(a_6^c,c_3)$ so to
improve the EOB/NR agreement further. Some subdominant multipoles
are rather robust in the nonspinning case, see e.g. Fig.~\ref{fig:hm}.
For large spins, we find the same problems related to the correct
determination of NQC corrections found for \TEOBResumShm{}~\cite{Nagar:2020pcj}.
Highly-accurate NR simulations covering a larger portion
of the parameter space (see e.g.~Ref.~\cite{Ramos-Buades:2019uvh})
are thus needed to robustly validate the model when
$e_\omega^{\rm NR}\gtrsim 0.2$.

\section{Conclusions}
We illustrated that minimal modifications to \TEOBResumShm{}~\cite{Nagar:2020pcj}
enabled us to build a (mildly) eccentric waveform model
that is reasonably NR-faithful over a nonnegligible portion of
the parameter space. This model could provide new eccentricity
measurements on LIGO-Virgo events. Our approach can be applied 
also in the presence of tidal effects. Higher-order corrections in the
waveforms and flux (see Refs.~\cite{Mishra:2015bqa,Cao:2017ndf,Hinderer:2017jcs})
should be included to improve the model for larger eccentricities.
In this respect, with a straighforward modification of the initial 
conditions~\cite{Damour:2014afa}, our model can also generate 
waveforms for dynamical captures or hyperbolic encounters~\cite{East:2012xq}, 
although NR validation is needed~\cite{Gold:2012tk,Damour:2014afa}. 
Provided high-order,  gravitational-self-force informed, resummed expressions 
for the EOB potentials~\cite{Akcay:2012ea,Bini:2013rfa,Akcay:2015pjz,Antonelli:2019fmq,Barack:2019agd},
as well as analytically improved fluxes to enhance the analytical/numerical
agreement of Fig.~\ref{fig:geo} for larger eccentricities,
we believe that our approach can pave the way
to the efficient construction of EOB-based waveform 
templates for extreme mass ratio inspirals, as interesting
sources for LISA~\cite{Babak:2014kqa,Berry:2019wgg}.

\begin{acknowledgments}
We are grateful to T.~Damour, G.~Pratten  and I.~Romero-Shaw for useful comments,
and to P.~Rettegno for cross checking many calculations.
\end{acknowledgments}
\bibliography{refs20201028.bib,local.bib}

\end{document}